\documentstyle{mn}
\input{psfig}

\newcommand{\otwo}{[O{\sc ii}]\ }

\title{The complex gravitational lens system B1933+503}

\author[Sykes et al.]
{C.~M. Sykes$^1$,  I.~W.~A. Browne$^1$, N.J. Jackson$^1$, D.R. Marlow$^1$,\\
 S.~Nair$^1$,P.~N. Wilkinson$^1$, R.~D. Blandford$^2$, J. Cohen$^2$,\\
 C.~D. Fassnacht$^2$, D. Hogg$^2$, T.J. Pearson$^2$, A.~C.~S. Readhead$^2$, D.~S. Womble$^2$,\\
 S.~T. Myers$^{2,5}$, A.~G. de Bruyn$^{4,6}$, M. Bremer$^3$, G.~K. Miley$^3$ $\&$
R.~T. Schilizzi$^4$\\
$^{1}$University of Manchester, NRAL Jodrell Bank, Macclesfield, Cheshire
SK11 9DL, England\\
$^{2}$California Institute of Technology,
Pasadena, CA 91125, USA \\ 
$^{3}$Sterrewacht Leiden, Postbus 9513, 2300RA Leiden, Netherlands\\
$^{4}$Joint Institute for VLBI in Europe, Postbus
2, 7990 AA, Dwingeloo, Netherlands\\ 
$^{5}$Dept. of Physics and Astronomy
University of Pennsylvania 209 S. 33rd Street Philadelphia, PA
19104, USA\\
$^{6}$Kapteyn Laboratory, Postbus 800, 9700 AA
Groningen, Netherlands }

\date{Oct 17, 1997}

\begin{document}
\maketitle

\vskip 3mm
\begin{abstract}
We report the discovery of the most complex arcsec-scale radio
gravitational lens system yet known. B1933+503 was found during the
course of the CLASS survey and MERLIN and VLA radio maps reveal up to
10 components. Four of these are compact and have flat spectra; the
rest are more extended and have steep spectra. The background lensed
object appears to consist of a flat spectrum core (quadruply imaged)
and two compact ``lobes'' symmetrically disposed relative to the
core. One of the lobes is quadruply imaged while the other is doubly
imaged. An HST observation of the system with the WFPC2 shows a galaxy
with an axial ratio of 0.5, but none of the images of the background
object are detected. A redshift of 0.755 has been measured for the
lens galaxy.

\end{abstract}

 
\baselineskip 12pt

\section{Introduction}

The Cosmic Lens All-Sky Survey (CLASS) is a survey of flat-spectrum
radio sources whose primary purpose is the discovery of new radio-loud
gravitational lens systems. A survey of 10,000 radio sources is being
undertaken. Those sources which, when observed with the VLA at 8.4~GHz
at a resolution of 200~mas, are found to possess multiple components
or very complex structure are followed-up at higher resolution with
MERLIN and the VLBA. Higher resolution allows one to separate the
genuine lens systems, which consist of multiple flat-spectrum
components, from systems containing a flat-spectrum core and
steep-spectrum extended radio emission. The discovery of the lenses
B1600+434, B1608+656 and B0712+472 has already been announced (Jackson
et al. 1995; Myers et al. 1995; Jackson et al. 1997).

In this paper we present VLA, MERLIN and VLBA radio images of
B1933+503. We also show a WFPC2 I image of the field together with an
optical spectrum obtained with the Keck Telescope. A list of the
observations is given in Table 1.

\begin{table}
\caption{Observations of B1933+503}
\begin{tabular}{ccccc}
\hline
Telescope&Observing&Exposure &Frequency      & Resolution \\
         &date     &time     &or $\lambda$   &(arcsec)    \\
\hline
VLA&1994 Mar 2&30 s&8.4 GHz&0.2\\
MERLIN&1995 Jun 23&1 hr&5.0 GHz&0.04\\
VLA&1995 July 6&30 min&15 GHz&0.13\\
VLA&1995 July 6&30 min&8.4 GHz&0.2\\
VLA&1995 Aug 28&1 hr&15 GHz&0.13 \\
VLA&1995 Sept 2&5 min&15 GHz&0.13 \\
MERLIN&1995 Oct 27&18 hr&1.7 GHz&0.12\\
Keck&1995 Sept 29&50 min &407--911nm&-\\
HST&1995 Nov 11&800s&540 nm& 0.1\\
HST&1995 Nov 11&1000s&814 nm&0.1\\
VLBA&1995 Nov 12&40 min&5.0 GHz&0.0015\\
\hline
\end{tabular}
\end{table}

\section{Radio and optical observations}

The radio maps of B1933+503 are presented in Figure~1. The components
are labeled 1 to 8; in addition we identify a weak component (1a)
close to 1 (visible in the MERLIN 1.7~GHz map).  Component 2 is
extended and its morphology suggests that it consists of two merging
images (see below). The MERLIN 1.7~GHz map also shows weak emission
between component 2 and component 7. A comparison of the MERLIN
1.7~GHz map with the MERLIN 5~GHz map clearly shows that the different
components have different spectral behaviours. Note, in particular,
components 4 and 5; the latter is one of the brightest components at
1.7~GHz while the former dominates at the higher frequencies. In Table
2 we list the flux densities of the radio components. In Figure~2 we
have combined this information to obtain radio spectra for the seven
strongest components. We see that 1, 3, 4 and 6 all have flat spectra
while the rest are steep. As expected, the flat spectrum components
are the most compact. Components 1, 3 and 4 are all detected
in the 6~cm VLBA map and all have compact emission on a scale of
$\sim$1~mas. The other flat spectrum component 6 is also detected, but
appears to have a lower surface brightness than components 1, 3 \&
4. None of the steep spectrum components are detected.

\begin{figure*}
\caption{Radio images of the B1933+503 system. {\bf Top left:} MERLIN
1.7-GHz image restored with a 120 $\times$ 120~mas. The contours are -1,
1, 2, 4, 8, 16, 32, and 64\% of the peak brightness of 0.017
Jy per beam. {\bf Top right:} VLA 15-GHz image restored with a 130 $\times$
130~mas beam. The contour levels are -4, 4, 8, 16, 32, 64\% of the
peak brightness of 0.018 Jy per beam. {\bf Bottom left:} MERLIN 5-GHz image
restored with a beam of 40 $\times$ 40~mas. The contour levels are -1.5,
1.5, 3, 6, 12, 24, 48, 96\% of the peak brightness of 0.019 Jy per
beam. {\bf Bottom right:} VLBA 5~GHz image restored with a beam of 10 $\times$
10~mas. The contour levels are -2, 2, 4, 8, 16, 32 and 64\% of the
peak brightness of 0.021 Jy per beam.}
\begin{tabular}{cc}
\psfig{file=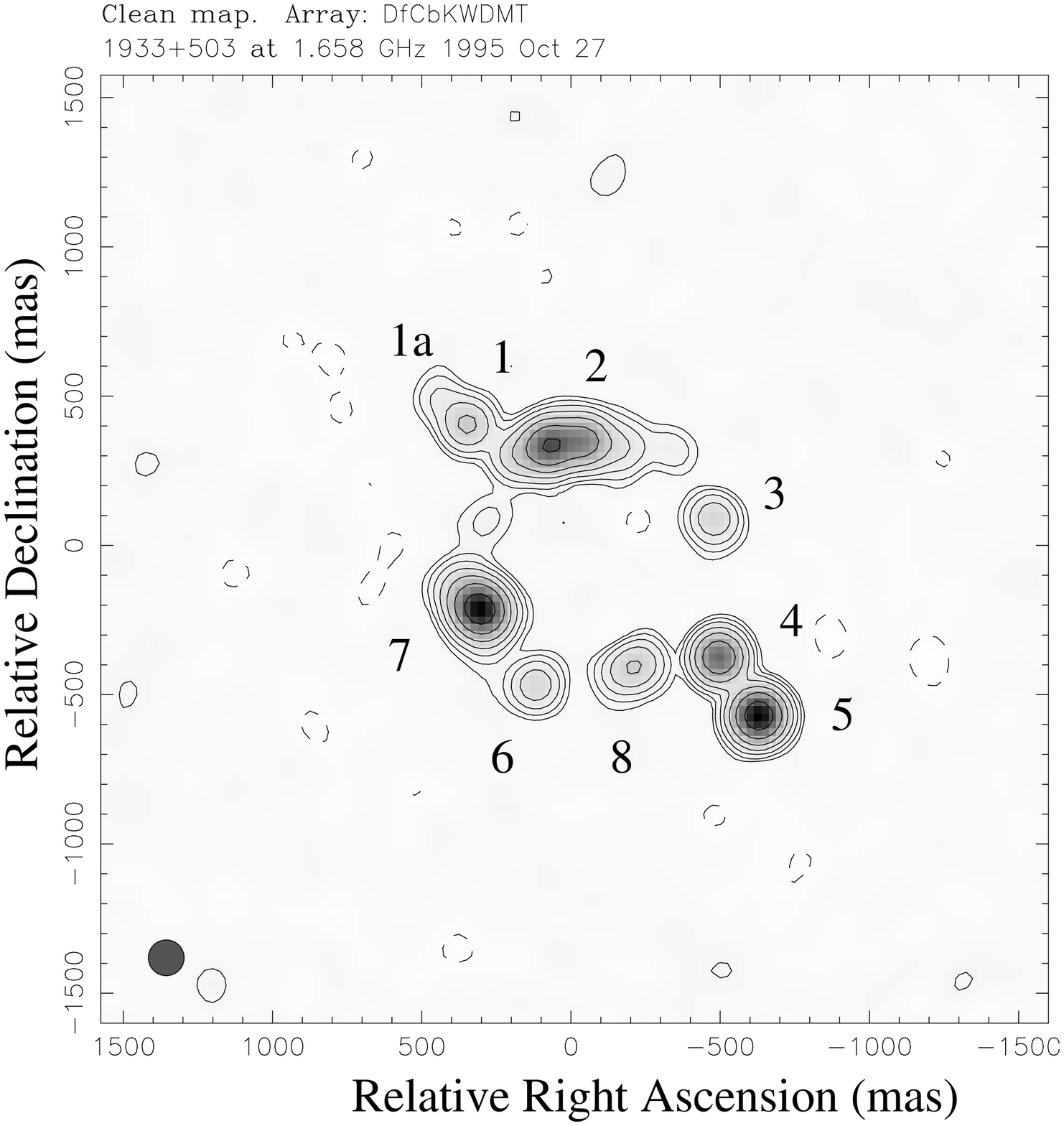,width=5.6cm,angle=0}&
\psfig{file=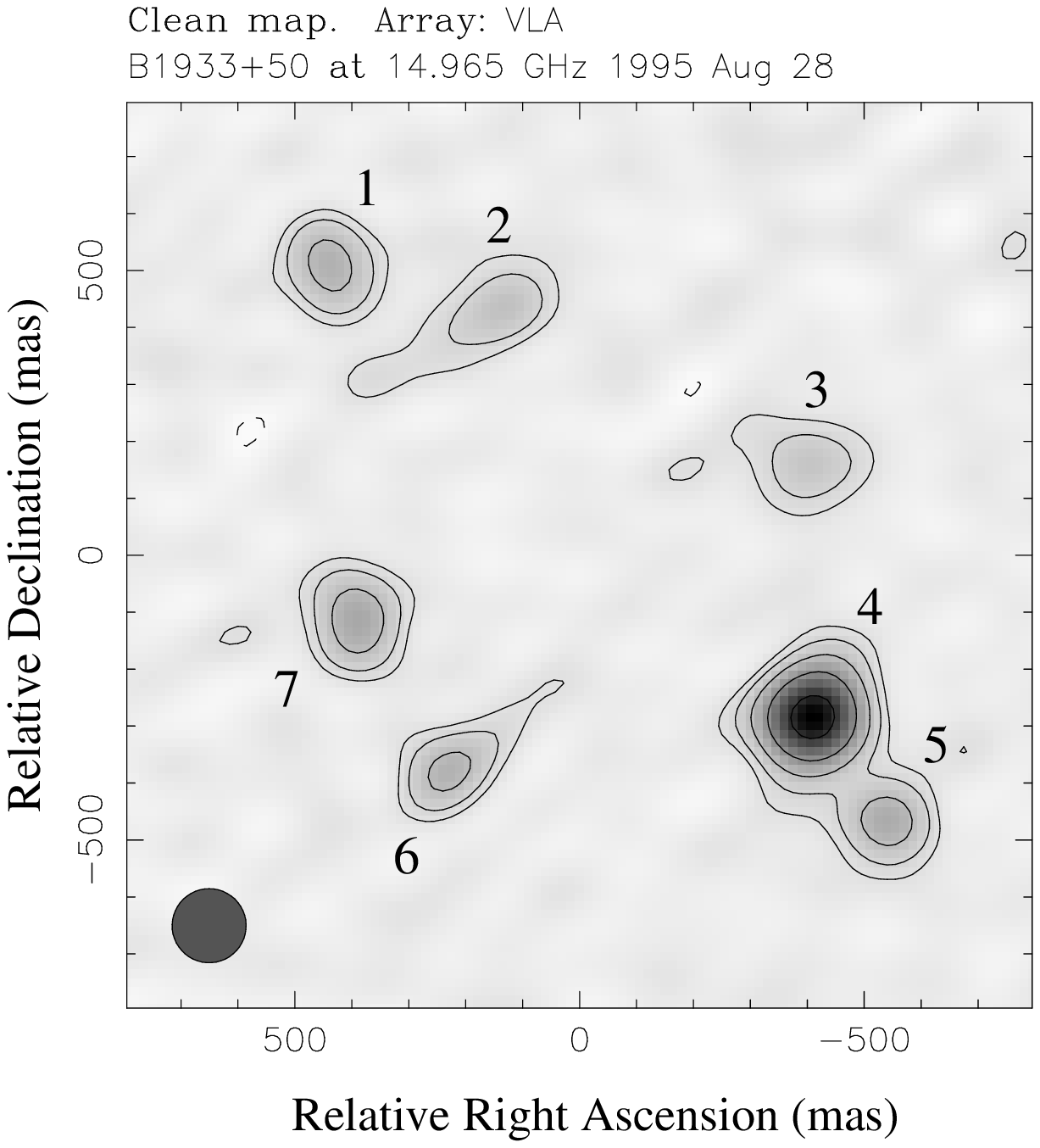,width=10cm,angle=-90}\\
\psfig{file=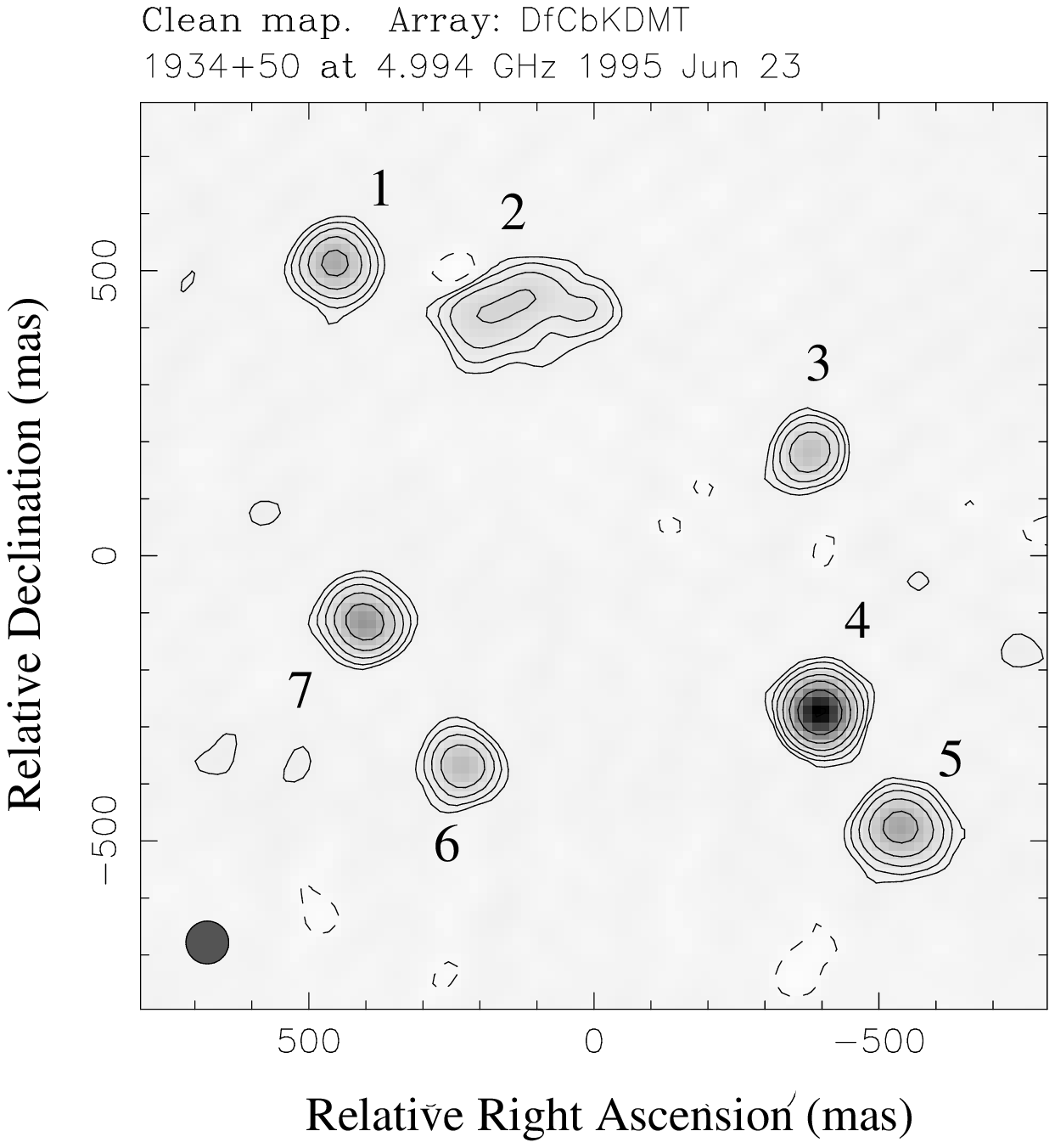,width=10cm,angle=-90}&
\psfig{file=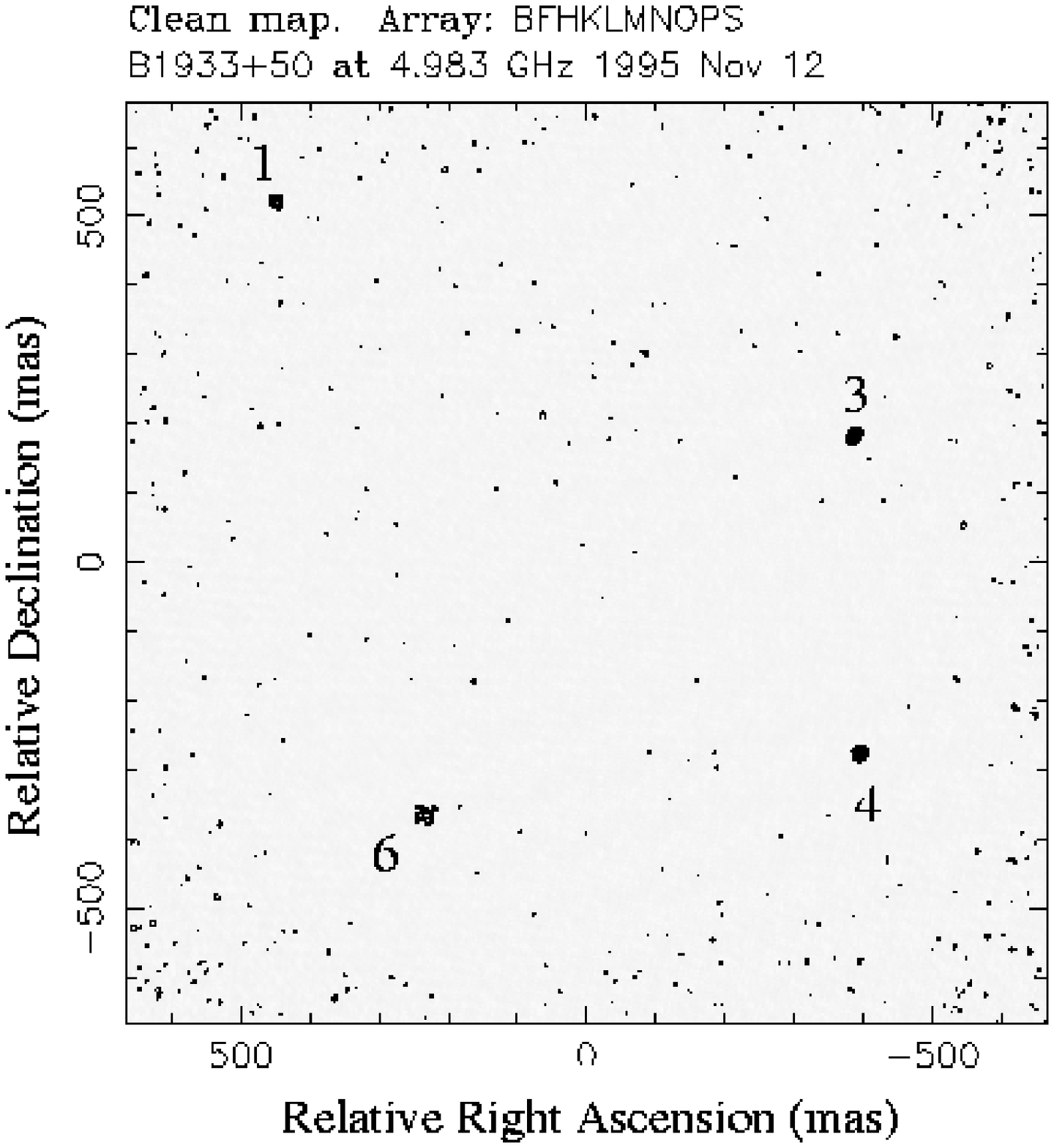,width=5.6cm,angle=-90}\\
\end{tabular}
\end{figure*}

\begin{figure*}
\caption{Radio spectra for the seven strongest components
of B1933+503.}
\psfig{file=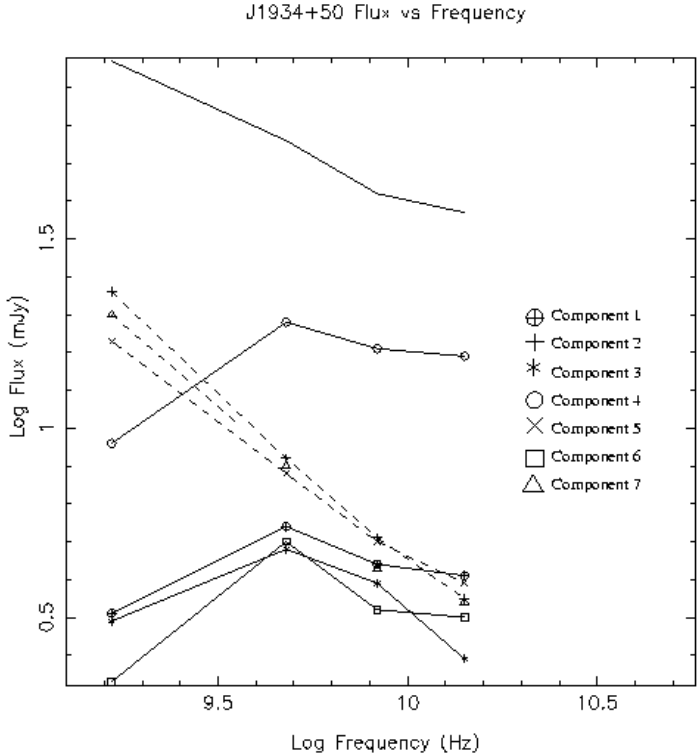,width=10cm,angle=-90}
\end{figure*}

\begin{figure*}
\caption{HST image taken with F814W filter}
\psfig{file=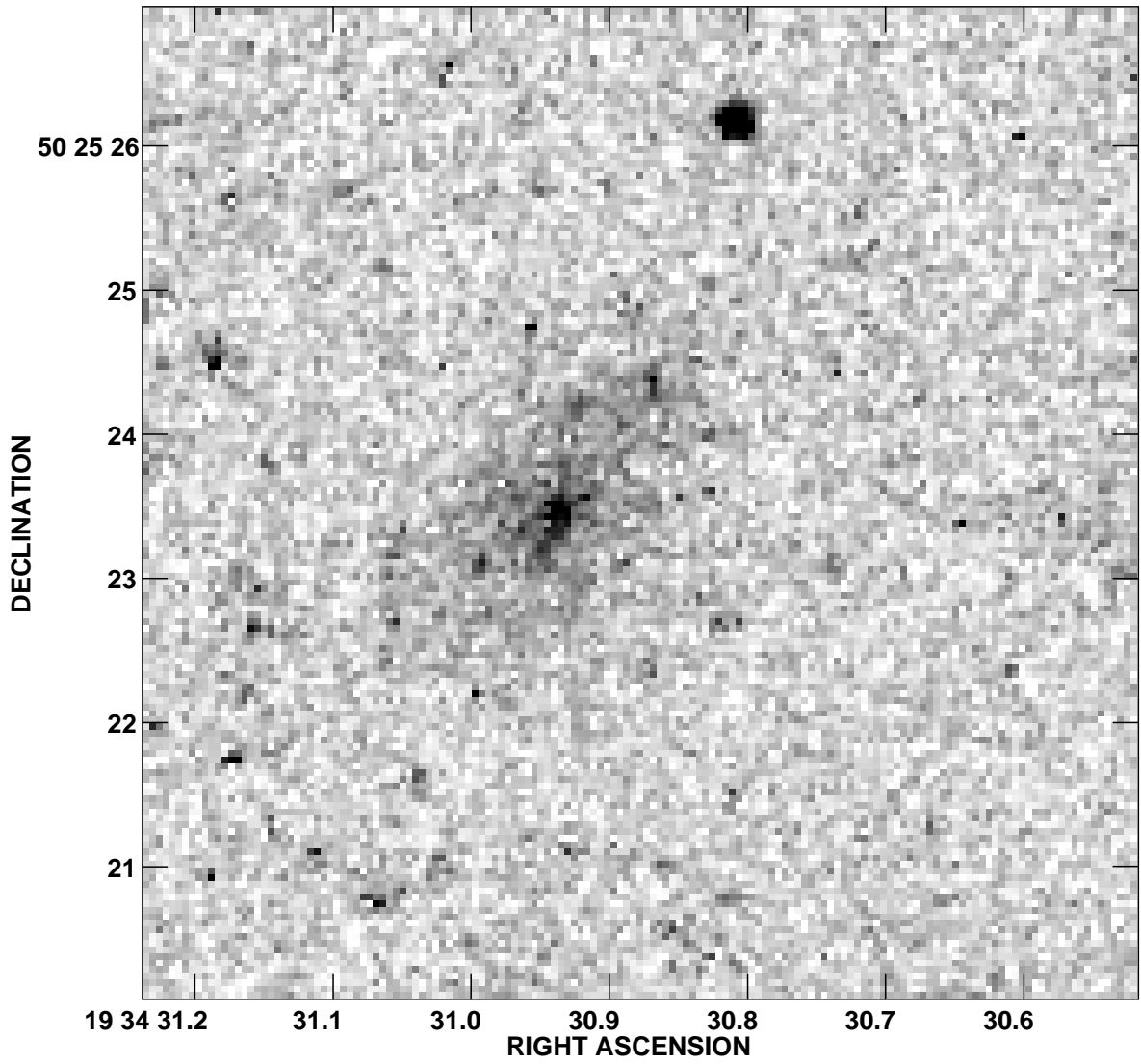,width=16cm,angle=270}
\end{figure*}

\begin{figure*}
\caption{Fits to the radial brightness distribution of the I-band
(814-nm) HST image. The data points represent the elliptical light
distribution collapsed on to the galaxy major axies. Each pixel is
gridded to the nearest 10 mas. The counts from any one pixel
contribute to just one point. Both disk (exponential) and de
Vaucouleurs  profiles fits  are shown. The latter gives the better fit if
the central part of the galaxy is included; a disk profile fits the
outer part of the galaxy well. The histogram shows the result of a cut
along the major axis.}
\psfig{file=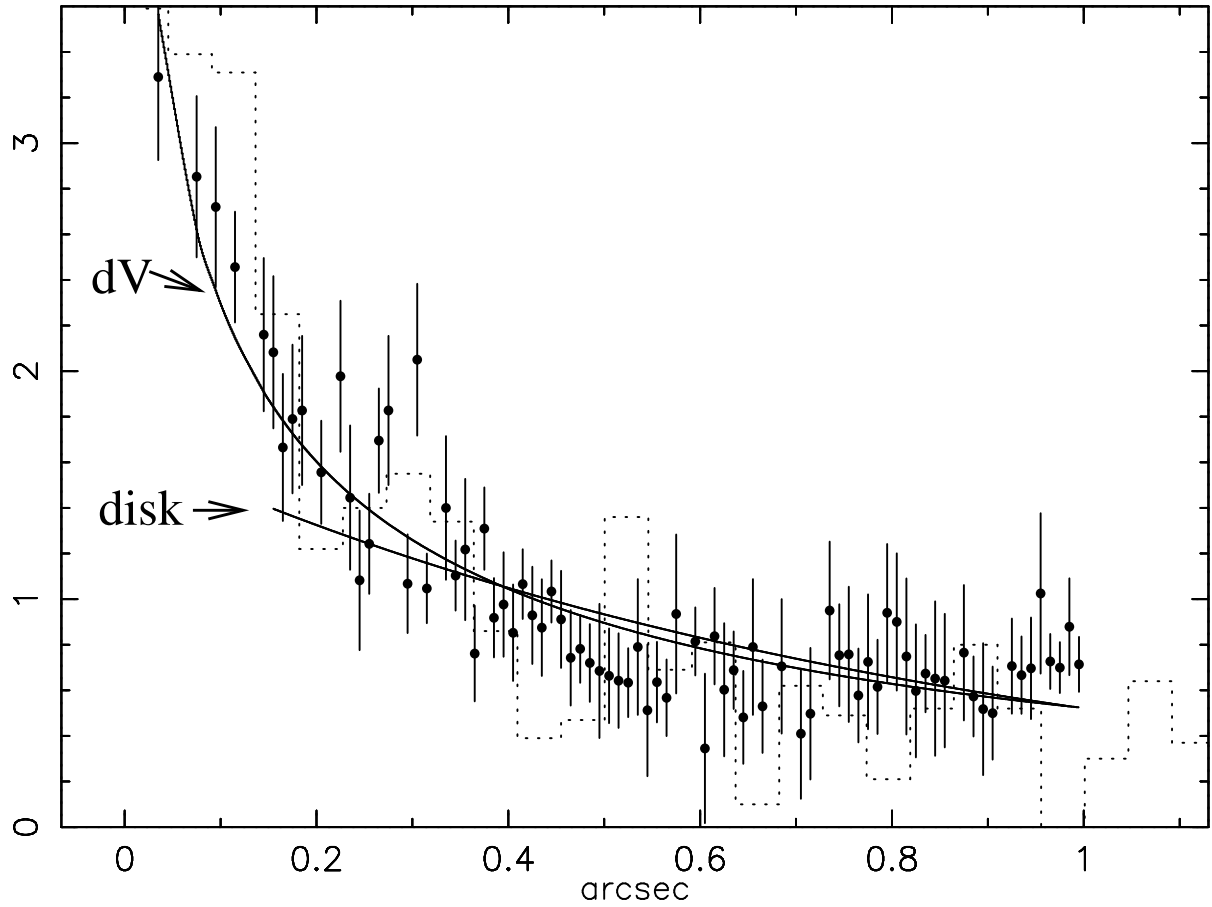,width=16cm,angle=-90}
\end{figure*}

\begin{figure*}
\caption{Keck spectrum of 1933+503 taken with
LRIS. Only that part of the spectrum between 6000\AA ~and 8000\AA ~is
displayed. The lower trace shows the rms noise on the spectrum.}
\psfig{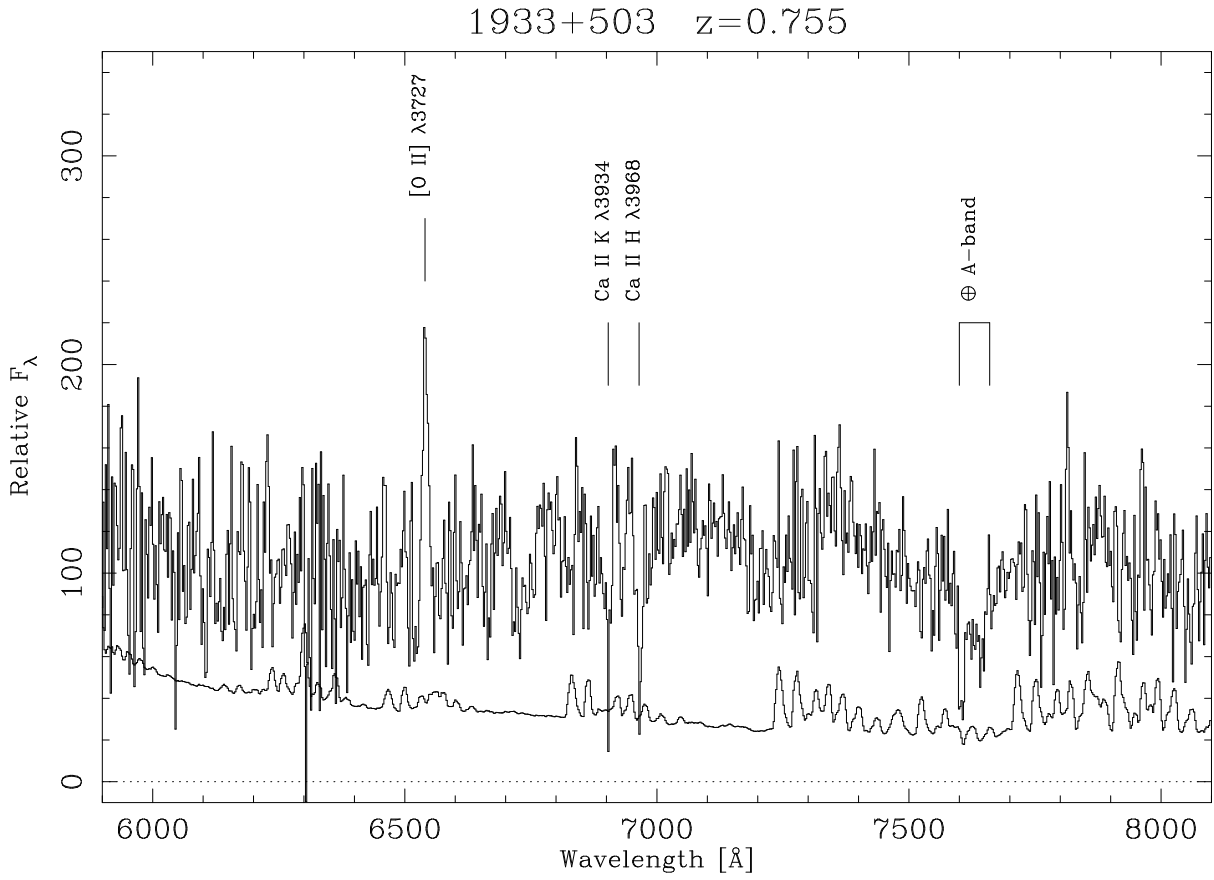}
\end{figure*}

In order to search for evidence of component variability we have
compared the three 15~GHz observations taken at different times. We
have used the analysis package DIFMAP (Shepherd et al. 1994) to fit a
model consisting of 7 elliptical gaussian components to the
data since components 1a and 8 are very weak at 15~GHz. There is some
evidence for variability; in particular the flat spectrum components
have all changed relative to the steep spectrum components between
1995 July 6 and Aug 28/Sept 2. Since the absolute flux density
calibration of the 15~GHz data is not good we have assumed that the
steep spectrum emission is not variable and have normalized each epoch
to that of Aug 28 in such a way that the sum of the steep spectrum
emission is the same on each occasion. In this way we deduce that the
flat spectrum components were systematically weaker by between 12\%
and 33\% in July 1995 as compared to Aug/Sept 1995. We take this as
tentative evidence for variability.

There is no optical emission from B1933+503 visible on the Palomar Sky
Survey. HST observations with WFPC2 were obtained on November 11, 1995
in both V and I bands. The HST observation, with a total exposure time
of 1100s in the F814W filter, shows a faint galaxy with a compact core
(Figure 3). We believe this to be the lensing galaxy. The integrated I
magnitude within an ellipse 2 $\times$ 1 arcsec is 20.6$\pm$0.2. The
galaxy position angle, derived by fitting elliptical isophotes to the
light profile, is $-$40$\pm$5 degrees. It has a bright compact core
with the ellipticity of the low brightness emission increasing to 0.5
outside the central 0\farcs6. In Figure 4 we show a fit to the radial
brightness distribution of the I-band (814-nm) HST image. The radial
brightness distribution has been obtained by assuming the galaxy to be
elliptical with axial ratio 0.5 and collapsing the data on to the
major axis.  Both disk and de Vaucouleurs models were fitted to the
profiles. Overall the de Vaucouleurs profile is the better fit
(reduced $\chi^{2}$ $\sim$1) though, if the central 0.2 arcsec is
ignored, the disk and de Vaucouleurs models are statistically
indistinguishable.  The radial profile favours classification as an
early-type galaxy, while the ellipticity of 0.5 and the [OII]3727
emission line (see below) are more typical of S0/spiral
systems. Clearly data with better sensitivity are needed for a more
definitive statement about the nature of the lensing galaxy.  The
galaxy is not clearly seen in the HST V image, leading to a limit on
its integrated V magnitude of about 22.5.

The images of the lensed object are not detected in the HST pictures, down to
limiting magnitudes of about 24.2 in I. If the lensed object is a quasar it
must therefore be extremely underluminous (especially as it would be
expected to be magnified by the lensing), or heavily reddened at
optical wavelengths by passage of the light through the lensing galaxy. 
The only possible point source seen in the image is about 1\farcs4 W,
2\farcs5 N of the lensing galaxy; however, no other images are seen
which might correspond to other radio components and we do not have
accurate astrometry which would enable us to locate it precisely on
the radio map.

An optical spectrum of B1933+503 was taken with the Keck telescope on
Sept 29 1995, using LRIS (Oke et al, 1995). The resulting spectrum is
shown in Figure~5.  Though the observing conditions were not ideal,
one narrow emission line and two absorption lines are detected. We
identify the emission line with \otwo at a redshift of 0.755 and the
absorption lines with CaII K \& H absorption with the same
redshift. The galaxy dominates the HST pictures, and hence the
emission in the spectrum must be nearly all light from the galaxy,
rather than from the images of the background object.

\section{B1933+503 as a lensed system}

A detailed discussion of a lens model for B1933+503 is given in the
companion paper (Nair, 1997). Here we simply discuss the observational
constraints in terms of a generic lens model. The large number of
features in the radio maps strongly suggests that a multi-component
background source is being imaged. We can use the radio spectral
information (Figure 2) and the surface brightness information from the
radio maps to identify features which could be images of a single
background component. It is evident that components 1, 3, 4, and 6 all
have very similar peaked spectra and are compact, though we note that
the VLBA map indicates that component 6 is somewhat less compact than
components 1, 3 and 4. These four components are almost certainly the
quadruple images of the core of the background radio
source. Components 2, 5 and 7 all have steep spectra down to
1.7~GHz. While the data on components 8 and 1a are less convincing it
is likely that they too have steep spectra. The above constraints have
been used as a starting point for the model presented by Nair
(1997). B1933+503 can be interpreted as the imaging by a single
elliptical lens of a compact triple source, consisting of an inverted
spectrum core and two steep spectrum ``lobes'', each separated by
$\sim$70~mas from the core.

B1933+503 is the first example of a gravitational lens where three
compact components are each multiply imaged.  As such, it offers the
prospect of deriving a quite detailed model of the surface density
distribution of the inner parts of the intervening galaxy. Measuring
relative time delays between image pairs will contribute to the
specificity of this model.  However, the failure to detect the lensed
images optically and, consequently, the poor prospects for
measuring the source redshift, makes us pessimistic that this source
will be useful for determining the Hubble constant.  More detailed
imaging studies with MERLIN and VLBA are underway.

\begin{table}
\begin{tabular}{ccccccc}
\hline
Cpt & \multicolumn{2}{c}{Positions (mas)} &
          \multicolumn{4}{c}{Flux Density (mJy)} \\
          & RA   & DEC  & 1.7GHz & 5GHz  & 8.4GHz & 15GHz \\
\hline
1         & 843.5  & 793.9  & 3.6   & 5.6  & 4.3   & 4.1  \\
2         & 519    & 720    & 23.0  & 8.3  & 4.4   & 3.5  \\
3         & 8.0    & 457.0  & 2.5   & 4.7  & 3.4   & 2.5  \\
4         & 0.0    & 0.0    & 9.4   & 19.4 & 15.7  & 15.5 \\
5         & -134   & -198   & 16.2  & 7.1  & 4.5   & 3.9  \\
6         & 627    & -88    & 2.2   & 5.4  & 3.6   & 3.2  \\
7         & 795    & 165    & 20.3  & 8.2  & 5.2   & 4.4  \\
8         & 283    & -36    & 3.6   & -    & 0.5   & -    \\
1a        & 942    & 883    & 0.9   & -    & -     & -    \\
\hline
\end{tabular}
\caption{Flux densities of components at various frequencies and their
positions measured relative to component 4 (Position (J2000) RA 19 34
30.899, DEC 50 25 23.22). Typical flux density errors are
$\pm$0.4~mJy. The errors on the relative positions vary; very
approximately those for components 1, 3 and 4 which have been obtained
from the VLBA 5~GHz data are $\leq$0.5~mas, for components 5, 6 and 7
from the MERLIN 5~GHz data are $\sim$5~mas and for components 8, 2 and
1a obtained from the MERLIN 1.7~GHz data they are $\sim$20~mas}
\end{table}




\section*{Acknowledgments}
 
This research used observations with the Hubble Space Telescope,
obtained at the Space Telescope Science Institute, which is operated
by Associated Universities for Research in Astronomy Inc. under NASA
contract NAS5-26555.  The Very Large Array is operated by Associated
Universities for Research in Astronomy Inc. on behalf of the National
Science Foundation. We thank the W.M. Keck foundation for the generous
grant that made the W.M. Keck Observatory possible.  MERLIN is
operated as a National Facility by NRAL, University of Manchester, on
behalf of the UK Particle Physics \& Astronomy Research Council.  This
work was supported in part by the US National Science Foundation under
grant AST-9420018 and in part by the European Commission, TMR
Programme, Research Network Contract ERBFMRXCT96-0034 "CERES".. CMS
and DRM have been supported by PPARC studentships.

\end{document}